# The transport character of quantum state in one-dimensional coupled-cavity-arrays: effect of the number of photons and entanglement degree

Shao-Qiang Ma (马少强)[1], Guo-Feng Zhang (张国锋)[1,2,3,*]

[1]*Key Laboratory of Micro-Nano Measurement-Manipulation and Physics (Ministry of Education), School of Physics and Nuclear Energy Engineering; State Key Laboratory of Software Development Environment, Beihang University, Xueyuan Road No. 37, Beijing 100191, China*

[2]*State Key Laboratory of Low-Dimensional Quantum Physics, Tsinghua University, Beijing 100084, China*

[3]*Key Laboratory of Quantum Information, University of Science and Technology of China, Chinese Academy of Sciences, Hefei 230026, China*

**Abstract**: The transport properties of the photons injected into one-dimensional coupled-cavity-arrays (CCAs) are studied. It is found that the number of photons cannot change the evolution cycle of the system and the time points at which W states and NOON state are obtained with a relatively higher probability. Transport dynamics in the CCAs exhibits that entanglement-enhanced state transmission is more effective phenomenon, and we show that for a quantum state with the maximum concurrence, it can be transmitted completely without considering the case of photon loss.

# I. Introduction

In the past few years, much attention has been paid to a coupled-cavity-arrays (CCAs) system，which is an effective platform to realize quantum information processing and to carry out photonics tasks due to their extremely rich physical properties as well as their wide potential applications [1-10]. For example, CCAs is known as an attractive controllable test bed for quantum simulators of many-body physics and strongly correlated many-body models [1-5, 11]. Transport properties of photons and various excitations in CCAs embedded with natural or artificial atoms are also receiving considerable attention [12-23]. In particular, it has been demonstrated that CCAs with a cyclic three-level system or a single atom open the possibility to quantum routing of single photon [24, 25].

---

[*] Correspondence and requests for materials should be addressed to G.Z.(gf1978zhang@buaa.edu.cn)

In this paper, we will study the properties of the transmission cycle of the photons injected into CCAs, the generation of W state and NOON state and the transport properties of entangled state in CCAs. Firstly we obtain the relationship between transport cycle of the photons in CCAs and the number of optical cavities. Secondly, for a given initial state, the state of the system at any time can be calculated. Therefore, we can choose a specific initial state so that we can obtain the W state and NOON state with a high probability at a certain moment. Finally, we demonstrate that the CCAs can realize the complete transmission of entangled state at a certain time by assuming that the two cavities at the beginning of CCAs are initially in an entangled state like $|\psi_{12}\rangle = \sin\theta|10\rangle + \cos\theta|01\rangle$.

The rest of this paper is organized as follows. In Sec II, the physical model and its solution are introduced. In Sec III, we will analyze the influence of the number of optical cavity and photon on transfer cycle of the photons injected into CCAs. W state and NOON state are prepared with a relatively higher possibility at a certain moment by regulating the system parameter when considering a weak coherent state as the initial one in Sec IV. In Sec V, we analyze transport properties of entangled state and show that the transmission of state is more effective with the increase of the entanglement degree even considering the presence of photon loss. Moreover, a complete transmission can be realized at a certain moment for a maximally entangled initial state without considering the case of photon loss. Finally, conclusions and discussions are given.

## II. Model and the solutions

Let's take a finite length CCAs in which nearest-neighbor cavity couples together so that it can realize photon transition. We hypothesized that all the cavities are identical, and each cavity sustains a single mode field (frequency $\omega$). The Hamiltonian of the system ($\hbar = 1$) is written as

$$H = \omega \sum_{j=1}^{N} a_j^+ a_j + J \sum_{j=1}^{N-1}(a_j^+ a_{j+1} + a_{j+1}^+ a_j), \tag{1}$$

where $a_j^+(a_j)$ is a photonic creation (annihilation) operator at the $j-th$ cavity. The second part in Eq. (1) represents the hopping interaction between inter-cavitys with coupling strength $J$. The Hamiltonian can be diagonalized by Jordan-Wigner transformation [26,27]

$$c_k = \sum_{j=1}^{N} a_j S(j,k), \tag{2}$$

with the following inverse transformation

$$a_j = \sum_{k=1}^{N} c_k S(j,k), \tag{3}$$

and the transformation matrix is defined by

$$S(j,k) = \sqrt{\tfrac{2}{N+1}} \sin(\tfrac{kj\pi}{N+1}). \tag{4}$$

In terms of Eqs. (2)-(4), the diagonalized Hamiltonian can be obtained

$$H = \sum_{k=1}^{N} \Omega_k c_k^+ c_k, \tag{5}$$

where $\Omega_k = \omega + 2J\cos[k\pi/(N+1)]$. $c_k^+(c_k)$ can also be considered as creation (annihilation) operator of the $k-th$ cavity associated $c$ mode. We assume that $|ck\rangle = |000...n_{ck}...0\rangle_c$ is the energy eigenstate with corresponding energy eigenvalue $n_{ck}\Omega_k$. $|ck\rangle$ is defined by $|ck\rangle = 1/\sqrt{n_{ck}!}\ (c_k^+)^{n_{ck}}|000...0...0\rangle$, where $|000...0...0\rangle$ is vacuum state. Similarly, $|aj\rangle = |000...m_{aj}...0\rangle$ is the eigenstate of $a_j^+ a_j$, $|aj\rangle$ is defined by $|aj\rangle = 1/\sqrt{m_{aj}!}\ (a_j^+)^{m_{aj}}|000...0...0\rangle$.

Therefore, making use of Eq. (2), energy eigenstate $|ck\rangle$ can be expressed in the photonic number representation

$$|ck\rangle = \tfrac{1}{\sqrt{n_{ck}!}} (\sum_{j=1}^{N} a_j^+ S^*(j,k))^{n_{ck}}|000...0...0\rangle. \tag{6}$$

According to Eq. (5) and Eq. (6), we can obtain the exact state evolution with time for a given initial state in the photonic number representation.

## III. Evolution cycle

We say that the system is periodic if the system is completely back to the initial state, and the time of the process is defined as the cycle of the system. The probability of the occurrence of initial state is denoted by $P_{nm}(t)$ ($n\ m$ represent the $n$ cavities and $m$ photons respectively), obviously the cycle of $P_{nm}(t)$ is the same with the one of this system. For the convenience of calculation, we only study the cycle of $P_{nm}(t)$, and the relevant conclusions we get are also applicable to the system's cycle. Now consider the case that there are three cavities in the CCAs. For an initial state like $|\psi(0)\rangle = |100\rangle$ (here and throughout we take the photonic number representation), $P_{31}(t)$ is analytically obtained as $P_{31}(t) = \cos^4(Jt/\sqrt{2})$. Next we choose $|200\rangle$ as the initial state, namely there are two photons in the first cavity, and $P_{32}(t) = \cos^8(Jt/\sqrt{2})$ can be acquired. In contrast to the above two cases, we find that, for a given initial state $|m00\rangle$, the function $P_{3m}(t)$ has the same period which has nothing to do with $m$.

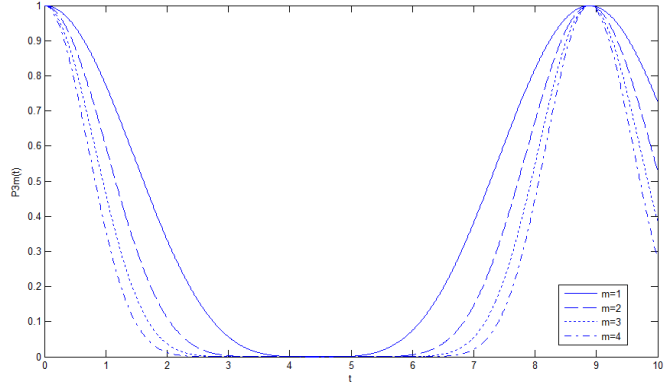

Fig.1: The evolution of $P_{3m}(t)$ with $t$ for the different photonic number in the initial state $|m00\rangle$ and here we take $J=0.5\omega$, $\omega=1$.

From Fig.1, one can find that $P_{3m}(t)$ has the same period (we denote it by $T$) for different photonic number. Further, we have $T = \sqrt{2}\pi/J$. In fact, these results can also be reflected in the process of solving the solution. We denote the number of the optical cavity by $n$. When $n = 3, m = 1, |\psi(0)\rangle = |100\rangle$, $|\psi(t)\rangle$ can be decomposed into the superposition of three eigenstates.

$$|\psi(t)\rangle = \tfrac{1}{2}|100\rangle_c e^{-i\Omega_{31}t} + \tfrac{\sqrt{2}}{2}|010\rangle_c e^{-i\Omega_{32}t} + \tfrac{1}{2}|001\rangle_c e^{-i\Omega_{33}t} \quad, \tag{7}$$

where energy eigenvalue $\Omega_{3k} = \omega + 2J\cos(k\pi/4)$. One can obtain

$P_{31}(t) = \quad \tfrac{1}{4} + \tfrac{1}{8}e^{i(\Omega_{31}-\Omega_{32})t} + \tfrac{1}{16}e^{i(\Omega_{31}-\Omega_{33})t} + \tfrac{1}{8}e^{-i(\Omega_{31}-\Omega_{32})t} + \tfrac{1}{8}e^{i(\Omega_{32}-\Omega_{33})t} +$

$\tfrac{1}{8}e^{-i(\Omega_{31}-\Omega_{33})t} + \quad \tfrac{1}{8}e^{-i(\Omega_{32}-\Omega_{33})t}$

(8)

Therefore the cycle of $P_{31}(t)$ is the least common multiple of $2\pi/|\Omega_{31} - \Omega_{32}|$, $2\pi/|\Omega_{31} - \Omega_{33}|$ and $2\pi/|\Omega_{32} - \Omega_{33}|$. The similar conclusion will be obtained regardless of the value of $m$ for $n=3$. Accordingly, we can prove that, for a given $n$, the system is periodic as long as the least common multiple of corresponding $2\pi/|\Omega_{ni} - \Omega_{nk}|$ ($i, k \in \{1,2,...n\}, \Omega_{nk} = \omega + 2J\cos[k\pi/(n+1)]$ exists. In addition, an interesting phenomenon can be found: when $m = 1, 2, 3$ and $4$, the corresponding $P_{3m}(t)$ are respectively $\cos^4(Jt/\sqrt{2})$, $\cos^8(Jt/\sqrt{2})$, $\cos^{12}(Jt/\sqrt{2})$ and $\cos^{16}(Jt/\sqrt{2})$, therefore, we can deduce that $P_{3m}(t) = \cos^{4m}(Jt/\sqrt{2})$. Let $m$ tends to infinity, time evolution of $P_{3m}(t)$ are shown in Fig.2. In fact, we can also obtain the results through the trend of $P_{3m}(t)$ with $m$ shown in Fig.1.

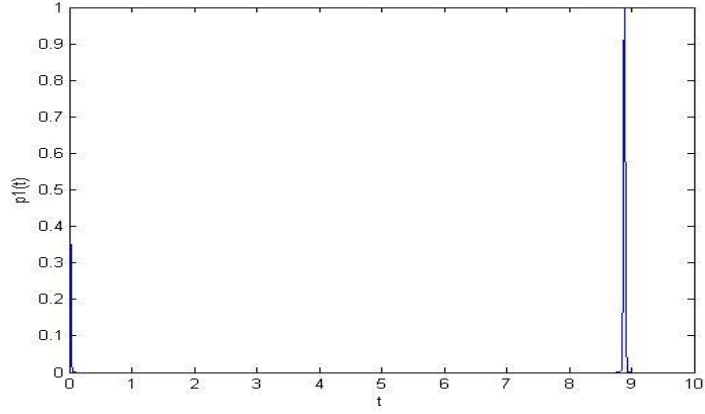

Fig.2: The evolution of $P_{3m}(t)$ for $m = 3000$ in the initial state $|m00\rangle$ and here we take $J=0.5\omega$, $\omega=1$.

## IV. Generation of W States and NOON States with a relatively higher probability

In this section, we take $J=0.5\omega$ and $\omega=1$. For $n=3$ case, we choose a weak coherent state as the initial state, i.e., $|\psi(0)\rangle = e^{-|\alpha_1|^2/2}(|0\rangle + \alpha_1|1\rangle) \otimes e^{-|\alpha_2|^2/2}(|0\rangle + \alpha_2|1\rangle) \otimes e^{-|\alpha_3|^2/2}(|0\rangle + \alpha_3|1\rangle)$, where $|\alpha_1| \ll 1$, $|\alpha_2| \ll 1$ and $|\alpha_3| \ll 1$. Thus $|\psi(t)\rangle = b_0(t)|000\rangle + b_{11}(t)|100\rangle + b_{12}(t)|010\rangle + b_{13}(t)|001\rangle + b_{2r}(t)|\psi_r\rangle$, in which $b_{2r}(t)|\psi_r\rangle$ represents the remainder, and it is obvious that $p_r(t) \ll p_{11}(t)$, $p_{12}(t), p_{13}(t) \ll p_0(t)$, ($p_{1i}(t) = |b_{1i}(t)|^2$, $p_r(t) = |b_r(t)|^2$, $p_0(t) = |b_0(t)|^2$). $p_0(t) = 1/[(|\alpha_1|^2 + 1)(|\alpha_2|^2 + 1)(|\alpha_3|^2 + 1)]$, that is to say $p_0(t)$ (the probability of occurrence of the vacuum state) does not change with the time, and has nothing to do with $J$ and $\omega$, moreover, $p_0(t)$ is close to 1. In this situation, the first and the third cavity enjoy the same status. If $\alpha_1 = \alpha_2$, we are more likely to get W states and NOON states. Consider the following three cases:

case1:　　$\alpha_1 = \alpha_2 = \alpha_3 = 0.1i$;
case2:　　$\alpha_1 = \alpha_3 = 0.01i$ and $\alpha_2 = 0.1i$;
case3:　　$\alpha_1 = \alpha_3 = 0.1i$ and $\alpha_2 = 0.01i$;

Time evolutions of the probabilities are shown in Fig.3.

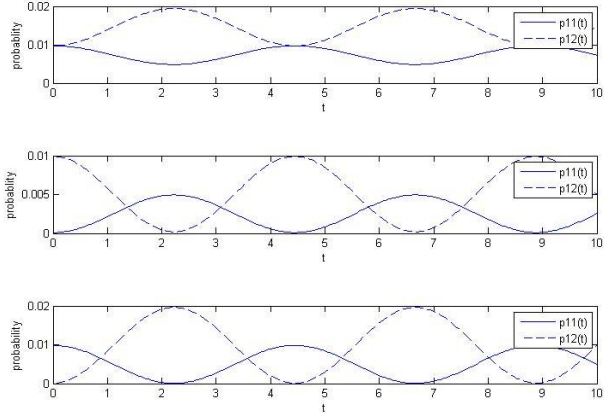

Fig.3: $p_{11}(t)$ (solid line) and $p_{12}(t)$ (dashed line). The three diagrams represent case1, case2 and case3, respectively (from top to down). $p_{11}(t) = p_{13}(t)$, since the first and the third cavities enjoy a coordinate position.

Similarly, $b_{2r}(t)|\psi_r\rangle = b_{21}(t)|200\rangle + b_{22}(t)|020\rangle + b_{23}(t)|002\rangle + b_{24}(t)|110\rangle + b_{25}(t)|011\rangle + b_{26}(t)|101\rangle + b_{3r}(t)|\psi_r\rangle$, in which $b_{3r}(t)|\psi_r\rangle$ is remainder, obviously $p_{3r}(t) \ll p_{21}(t), p_{22}(t), p_{23}(t), p_{24}(t), p_{24}(t), p_{26}(t)$. Time evolution of the probabilities are shown in Fig.4.

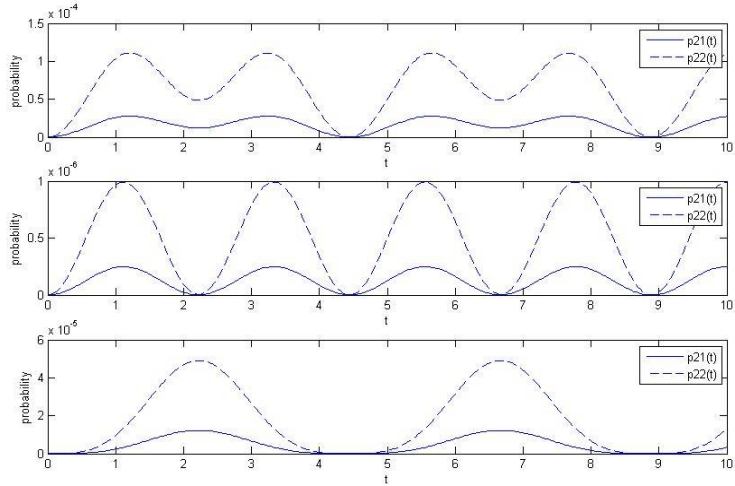

Fig.4: $p_{21}(t)$ (solid line) and $p_{22}(t)$ (dashed line). The three diagrams represent case1, case2 and case3 respectively (from top to down). $p_{21}(t) = p_{23}(t)$, since the first and the third cavity have the same status in the system.

Similar to the above two cases, we can get the following figure for the remainder $b_{3r}(t)|\psi_r\rangle$ ($p_{31}(t)$, $p_{32}(t)$ and $p_{33}(t)$ represent the probabilities of the occurrence of $|300\rangle$, $|030\rangle$ and $|003\rangle$).

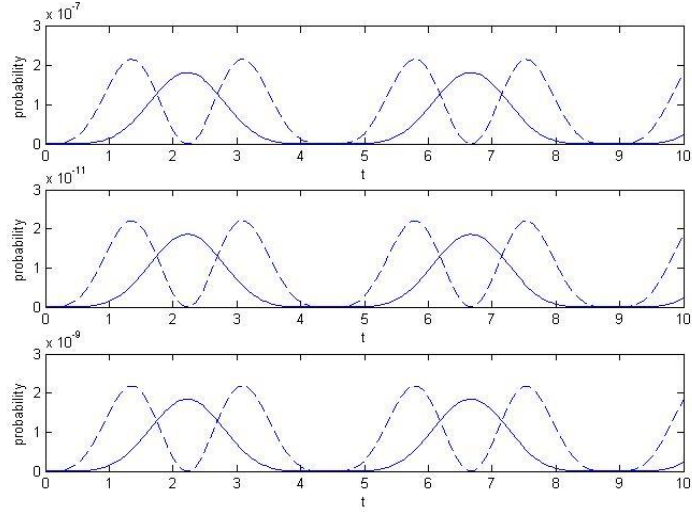

Fig.5: $p_{31}(t)$ (solid line) and $p_{32}(t)$ (dashed line). The three diagrams represent case1, case2 and case3 respectively (from top to down). $p_{31}(t) = p_{33}(t)$, since the first and the third cavity have the same status in the system.

In Fig.3 (and Fig.5), it is clear that there exist some cross points of the probabilities. At these special time, $p_{11}(t) = p_{12}(t) = p_{13}(t)$ ($p_{31}(t) = p_{32}(t) = p_{33}(t)$), which indicates the W states containing one (three) photon are generated with a relatively higher probability, and the corresponding time $t_w$ can be calculated. When $p_{11}(t)$ ($p_{31}(t)$) reaches the maximum value and $p_{12}(t)$ ($p_{32}(t)$) reaches the minimum value, NOON states containing one (three) photon are generated with a higher probability and we denote the time by $t_n$. In Table I, we give the time at which W states and NOON states are obtained with a higher probability. Simultaneously, from Fig.4, it can be obtained easily that W states and NOON states about two photons will not be generated.

Table I: Time when W and NOON states are obtained with a higher probability ($k = 0,1,2,3 ...$).

| case | case1 | case2 | case3 |
| --- | --- | --- | --- |
| single photon W state | $t_w = \sqrt{2}k\pi$ | $t_w = 1.3612 + \sqrt{2}k\pi$ | $t_w = 0.8679 + \sqrt{2}k\pi$ |
| single photon NOON state | none | $t_n = \pi/\sqrt{2} + \sqrt{2}k\pi$ | $t_n = \sqrt{2}k\pi$ |
| two photons W state | none | none | none |

| two photons NOON state | none | none | none |
| three photons W state | $t_w = 1.7408+\sqrt{2}k\pi$ | $t_w = 1.7408+\sqrt{2}k\pi$ | $t_w = 1.7408+\sqrt{2}k\pi$ |
| three photons NOON state | $t_n = \pi/\sqrt{2} + \sqrt{2}k\pi$ | $t_n = \pi/\sqrt{2} + \sqrt{2}k\pi$ | $t_n = \pi/\sqrt{2} + \sqrt{2}k\pi$ |

Comparing Fig.5 and Fig.6, we can find the behaviors of probabilities with time are the same except for the amplitudes.

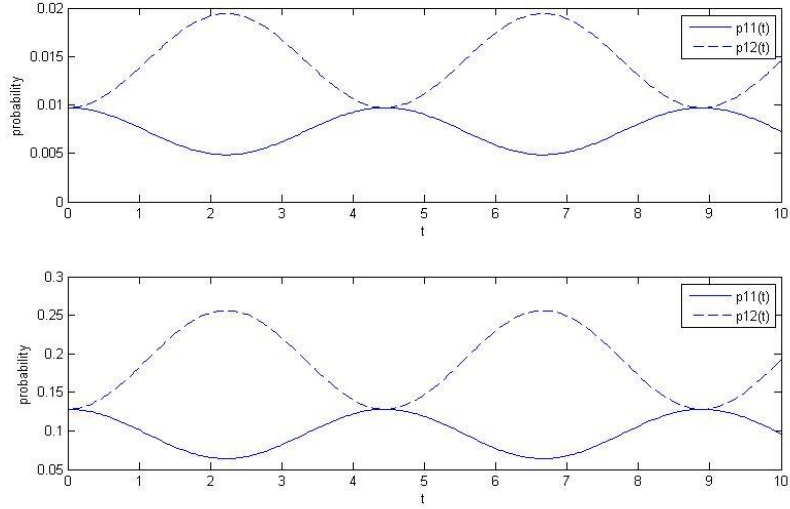

Fig.6: $p_{11}(t)$ (solid line) and $p_{12}(t)$ (dashed line), the two diagrams represent $\alpha_1 = \alpha_2 = \alpha_3 = 0.1i$ and $\alpha_1 = \alpha_2 = \alpha_3 = 0.5i$, respectively (from top to down).

As has been mentioned above, when $n = 3$, the first and the third cavity have the same status in the system. If let them also has the same weight in the initial state, then W states and NOON states are more likely to be obtained. Therefore, we can also choose $|0m0\rangle$ as an initial state. For the case $m=1$, the probability are analytically obtained as $p_1(t) = 0.5\sin^2(\sqrt{2}Jt)$, $p_2(t) = \cos^2(\sqrt{2}Jt)$ and $p_3(t) = 0.5\sin^2(\sqrt{2}Jt)$. Figure 7 gives a plot of the probabilities versus time.

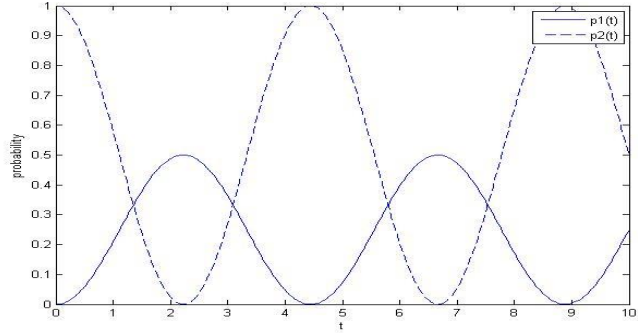

Fig.7: $p_1(t)$ (solid line) and $p_2(t)$ (dashed line). Here take $J = 0.5\omega$, and $\omega = 1$.

It is clear that there exist some cross points of the probabilities. At these special time, we have $p_1(t) = p_2(t) = p_3(t) = 1/3$, which indicates the W states are generated with hundred percent probabilities. Namely the time $t$ satisfies the equation $\sin^2(t/\sqrt{2}) = 2\cos^2(t/\sqrt{2})$. Similarly, the NOON states are generated if $t$ satisfies the equation $\cos^2(t/\sqrt{2}) = 0$. The solution of the equation is shown in Table II.

Table II: Time at which W and NOON states are completely obtained for the initial state $|010\rangle$, ($k = 0,1,2,3 ...$).

| W states generated time | NOON states generated time |
| --- | --- |
| $t_w = 1.3511 + \sqrt{2}k\pi$ or $t'_w = 3.0919 + \sqrt{2}k\pi$ | $t_n = \pi/\sqrt{2} + \sqrt{2}k\pi$ |

For $m=2$, we denote the probabilities of occurrence of $|200\rangle$, $|020\rangle$, $|002\rangle$, $|110\rangle$, $|011\rangle$ and $|101\rangle$ respectively by $p_1(t)$, $p_2(t)$, $p_3(t)$, $p_4(t)$, $p_5(t)$ and $p_6(t)$. Because of the symmetry of the system, it is obvious that $p_1(t) = p_3(t)$, and $p_4(t) = p_5(t)$. The results are illustrated in Fig.8.

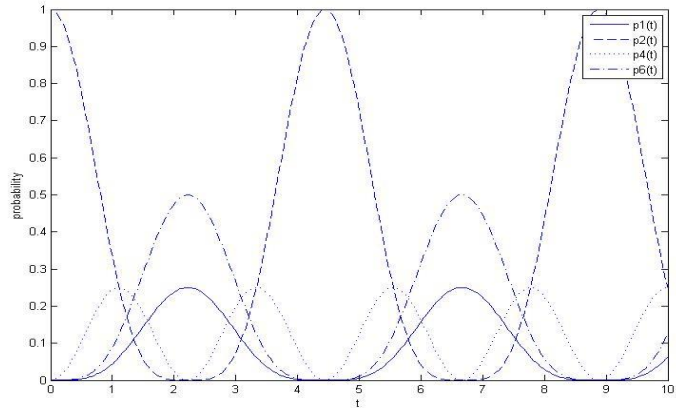

Fig.8: $p_1(t)$ (solid line), $p_2(t)$ (dashed line), $p_4(t)$ (dotted line) and $p_6(t)$ (dash-dotted line), here take $J = 0.5\omega$, and $\omega = 1$.

It is obvious that there also exist some cross points of $p_1(t)$ and $p_2(t)$, which indicates the W states are generated with a relatively higher possibility. That is to say, the time $t$ satisfies the equation $\sin^4(t/\sqrt{2}) = 4\cos^4(t/\sqrt{2})$. Similarly NOON states are generated if $t$ satisfies the equation $\cos^4(t/\sqrt{2}) = 0$. The detailed solution can be seen from Table III.

Table III: Time at which W and NOON states are obtained with a relatively higher possibility for the initial state $|020\rangle$, ($k = 0,1,2,3 ...$).

| W states generated time | NOON states generated time |
|---|---|
| $t_w = 1.3511 + \sqrt{2}k\pi$ or $t'_w = 3.0919 + \sqrt{2}k\pi$ | $t_n = \pi/\sqrt{2} + \sqrt{2}k\pi$ |

We find that when $t = t_w$ or $t'_w$, $p_3(t) = p_4(t)$. That is to say, when the W state is generated a relatively higher possibility, the probabilities of detecting of $|110\rangle, |011\rangle$ and $|101\rangle$ are equivalent. At time $t_n$, the probabilities of occurrence of $|200\rangle$ and $|200\rangle$ are equal to $1/4$, while the probability of occurrence of $|101\rangle$ is $1/2$. According to the two cases above ($m = 1, m = 2$), it is interesting to note that the moment that producing W states and NOON states has nothing to do with the number of photons in the CCAs.

## V. Effects of entanglement degree on the transmission of state

In this section, we presume that there are $n$ cavities in the CCAs, and we consider such a situation in which the photons are initially in the first two cavities with the following entangled state $|\psi_{12}\rangle = \sin\theta|1\rangle_1|0\rangle_2 + \cos\theta|0\rangle_1|1\rangle_2$. For the simplicity, we take $0 < \theta < \pi/2$ throughout the paper. The other cavities are in the CCAs are in vacuum state, and the initial entanglement can be measured by concurrence [28]

$$C(\theta) = |\sin 2\theta|, \quad (9)$$

Obviously, the concurrence vanishes when $\theta = 0$ and $\pi/2$, while the concurrence reaches the maximum value 1 when $\theta = \pi/4$.

If $n = 4$, then $|\psi(0)\rangle = \sin\theta|1000\rangle + \cos\theta|0100\rangle$. Denote the probability of detecting that entangled state transports from the first two cavities to the last two cavities by $p(t,\theta)$. One can obtain (here we take $J=0.5\omega, \omega=1$)

$$p(t,\theta) = \frac{1}{5}\sin^2\frac{t}{4}(20(\sin\theta\cos\theta)^2\cos^2\frac{\sqrt{5}t}{4} + 4\sin^2\frac{\sqrt{5}t}{4}), \quad (10)$$

According to Eq. (9), $p(t,\theta)$ can also be rewritten as following

$$p(t,C) = \frac{1}{5}\sin^2\frac{t}{4}(5C^2\cos^2\frac{\sqrt{5}t}{4} + 4\sin^2\frac{\sqrt{5}t}{4}), \quad (11)$$

in which $C$ represents the concurrence of initial state. The evolution of $p(t,C)$ is presented in Fig.9.

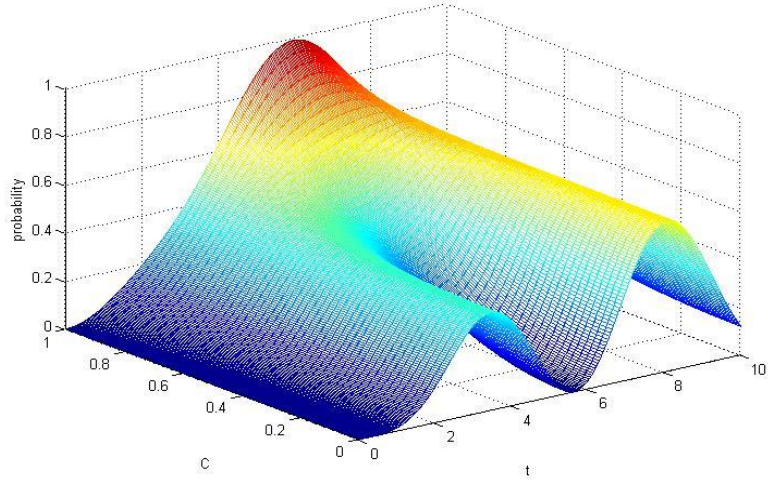

Fig.9: The probability of entangled state transports from the first two cavities to the last two cavities.

Calculating the partial derivative of $p(t,C)$ against $C$, one can acquire $\partial p/\partial C = 2C\cos^2(\sqrt{5}t/4)\sin^2(t/4)$. $p(t,C)$ is possible to achieve the maximum value, when $\partial p/\partial C = 0$ or $C = 1$ or $C = 0$. $\partial p/\partial C = 0$ means that $C = 0$ or $\cos^2(\sqrt{5}t/4) = 0$ or $\sin^2(t/4) = 0$, however, one can obtain that $p(t,C) \leq 0.8$. But $p(t,C) = 1$ while $C = 1$ and $t = 106.7957$. Therefore $p(t,C)$ will reach maximum value 1, only if $C = 1$. That is to say there exists a complete transmission of entangled state from the first two cavities to the last two cavities in the CCAs, when the concurrence of the initial state has the maximum value 1. Substituting $t = 106.7957$ into Eq. (10), one can obtain

$$p(C) = 0.9998715913626225(0.00008571750278695456 + C^2)$$    ,

(12)

The evolution of $p(C)$ is shown in Fig.10.

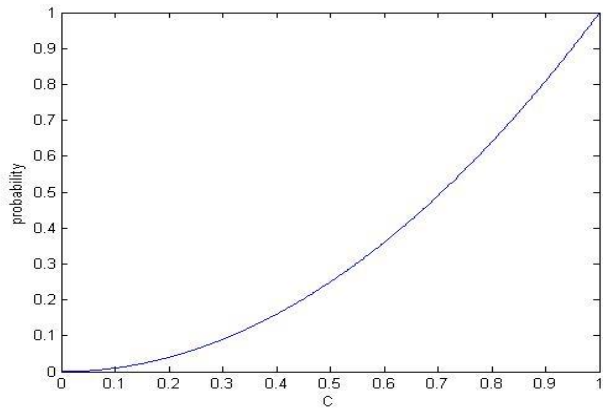

Fig.10: The evolution of $p(C)$ when $t = 106.7957$.

As we all know, the CCAs is not ideal in experiments, and then we consider a system consisting of dissipation. With the presence of photon loss, the time evolution of the whole system is described by the Lindblad master equation ($\hbar = 1$)

$$\frac{d\rho}{dt} = i[H,\rho] + \gamma \sum_{i=1}^{n} D(a_i)\rho, \tag{13}$$

where the superoperator is defined as $D(c)\rho = c\rho c^+ - 1/2(c^+c\rho + \rho c^+c)$. Here we have assumed that $n$ represents the number of cavity, $a_i$ is the annihilation operators of the $i-th$ cavity and the cavity mode $a_i$ is damped at rate $\gamma$.

We presume that there are three cavities in the CCAs, and we consider $|\Psi\rangle = |\psi_{12}\rangle \otimes |0\rangle_3 = [\sin\theta|1\rangle_1|0\rangle_2 + \cos\theta|0\rangle_1|1\rangle_2]|0\rangle_3$ as the initial state. Considering the presence of photon loss, we take $|0\rangle_1|0\rangle_2|0\rangle_3$, $|1\rangle_1|0\rangle_2|0\rangle_3$, $|0\rangle_1|1\rangle_2|0\rangle_3$ and $|0\rangle_1|0\rangle_2|1\rangle_3$ as base vectors by which the system can be fully described. Therefore the initial density is

$$\rho(0) = \begin{pmatrix} 0 & 0 & 0 & 0 \\ 0 & \sin^2\theta & \sin\theta\cos\theta & 0 \\ 0 & \sin\theta\cos\theta & \cos^2\theta & 0 \\ 0 & 0 & 0 & 0 \end{pmatrix} \tag{14}$$

Taking $J = 0.5\omega, \omega = 1$ and $\gamma = 0.1$, the evolved density matrix $\rho(t)$ of the system can be solved by numerical methods. We consider the state $|\Phi\rangle = |\psi_{23}\rangle \otimes |0\rangle_1 = \sin\theta|0\rangle_1|1\rangle_2|0\rangle_3 + \cos\theta|0\rangle_1|0\rangle_2|1\rangle_3$ as the final state. Denote $\langle\Phi|\rho(t)|\Phi\rangle$ by $p_{t\theta}$, and $p_{t\theta}$ indicates the probability of detecting that entangled state transports from the first two cavities to the last two cavities. The evolution of $p_{t\theta}$ versus $\theta$ at a certain time is presented in Fig. 11.

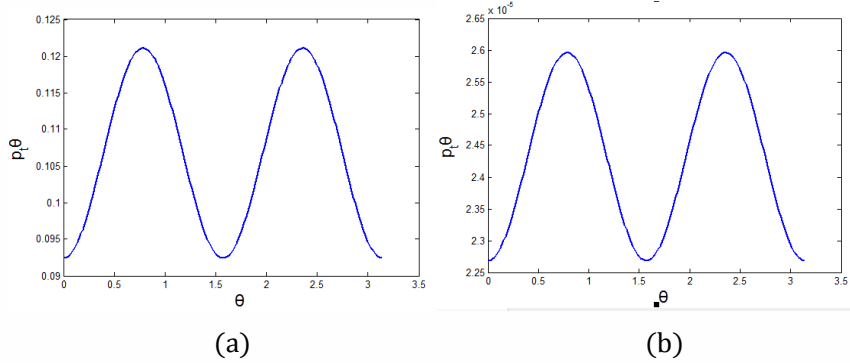

(a)　　　　　　　　　　　　　(b)

Fig.11: The probability of entangled state transports from the first two cavities to the last two cavities at the moment t=10. (a): the probability against $\theta$ at the moment t=100 (b): The probability against $\theta$ at the moment t=100.

From Fig.11, even taking the presence of dissipation into account, we can deduce that the probability of entangled state transports from the first two cavities to the last two cavities is proportional to the concurrence of initial state. The transmission probability is becoming smaller as time goes on due to the presence of photon loss.

**Conclusions**

In conclusion, we have investigated the properties of the photons injected into a one-dimensional coupled-cavity array (CCAs). First of all, the evolution period of the system is proved to be related to the number of the optical cavities. What's more, W states and NOON states can be obtained at certain moment by choosing symmetric state as initial one. The above two conclusions has nothing to do with the number of photons injected into the CCAs. In addition, it is shown that the probability of state transmission is proportional to the concurrence of initial state even considering the presence of photon loss and the maximal entangled initial state can be transmitted completely in the case that damping rate equal to zero.

**Acknowledgments**

This work is supported by the National Natural Science Foundation of China (Grant No. 11574022 and 11174024) and the Open Fund of IPOC (BUPT) grants Nos. IPOC2013B007, also supported by the Open Project Program of State Key Laboratory of Low-Dimensional Quantum Physics (Tsinghua University) grants Nos. KF201407, and Beijing Higher Education (Young Elite Teacher Project) YETP 1141 and the Fundamental Research Funds for the Central Universities of Beihang University (Grant No. YWF-15-WLXY-013).